\RequirePackage[mathlines]{lineno}
\documentclass[amsmath,amssymb,twocolumn,prd,floatfix,showpacs,nofootinbib]{revtex4-1}
\usepackage{tabularx}
\usepackage{bm, color} 
\usepackage{overpic,subfigure} 
\usepackage{multirow}
\usepackage{array}
\usepackage{dcolumn} 
\usepackage[symbol]{footmisc}
\usepackage{epstopdf}
\usepackage{ulem}
\usepackage{hyperref}

\RequirePackage{xspace}

\newcommand{\gev}{\ensuremath{\mathrm{\,Ge\kern -0.1em V}}\xspace}
\newcommand{\mev}{\ensuremath{\mathrm{\,Me\kern -0.1em V}}\xspace}
\newcommand{\mevcc}{\ensuremath{{\mathrm{\,Me\kern -0.1em V\!/}c^2}}\xspace}

\def\fz#1       {\ensuremath{f_0({#1})}\xspace}

\newcommand{\Dbf}{$(1.091\pm0.027\pm0.035)\%$}
\newcommand{\Dpbf}{$(1.135\pm0.034\pm0.031)\%$}

\begin{document}

\title{\boldmath Improved measurements of the absolute branching fractions of the inclusive decays $D^{+(0)}\to\phi X$}

\author{
M.~Ablikim$^{1}$, M.~N.~Achasov$^{10,d}$, P.~Adlarson$^{59}$, S. ~Ahmed$^{15}$, M.~Albrecht$^{4}$, M.~Alekseev$^{58A,58C}$, A.~Amoroso$^{58A,58C}$, F.~F.~An$^{1}$, Q.~An$^{55,43}$, Y.~Bai$^{42}$
% ...
\\
\vspace{0.2cm}
(BESIII Collaboration)\\
\vspace{0.2cm} {\it
$^{1}$ Institute of High Energy Physics, Beijing 100049, People's Republic of China\\
% ...
$^{4}$ Bochum Ruhr-University, D-44780 Bochum, Germany\\
% ...
$^{10}$ G.I. Budker Institute of Nuclear Physics SB RAS (BINP), Novosibirsk 630090, Russia\\
% ...
$^{15}$ Helmholtz Institute Mainz, Johann-Joachim-Becher-Weg 45, D-55099 Mainz, Germany\\
% ...
$^{42}$ Southeast University, Nanjing 211100, People's Republic of China\\
$^{43}$ State Key Laboratory of Particle Detection and Electronics, Beijing 100049, Hefei 230026, People's Republic of China\\
% ...
$^{55}$ University of Science and Technology of China, Hefei 230026, People's Republic of China\\
% ...
$^{58}$ (A)University of Turin, I-10125, Turin, Italy; (B)University of Eastern Piedmont, I-15121, Alessandria, Italy; (C)INFN, I-10125, Turin, Italy\\
$^{59}$ Uppsala University, Box 516, SE-75120 Uppsala, Sweden\\
% ...
$^{d}$ Also at the Novosibirsk State University, Novosibirsk, 630090, Russia\\
% ...
}
\vspace{0.4cm}
}

\begin{abstract}
By analyzing 2.93 fb$^{-1}$ of $e^+e^-$ annihilation data taken at the center-of-mass energy $\sqrt s=$ 3.773 GeV with the BESIII detector, we determine the branching fractions of the inclusive decays $D^+\to\phi X$ and $D^0\to\phi X$ to be {\Dpbf} and {\Dbf}, respectively, where $X$ denotes any possible particle combination. The first uncertainties are statistical and the second systematic. We also determine the branching fractions of the decays $D\to\phi X$ and their charge conjugate modes $\bar{D}\to\phi \bar{X}$ separately for the first time, and no significant CP asymmetry is observed.
\end{abstract}


\maketitle

\section{INTRODUCTION}

Experimental studies of the inclusive $D\to \phi X$ decays, where $X$ denotes any possible particle combination, are important for charm physics due to the following reasons.
Firstly, precise measurements of their branching fractions
offer an independent check on the existence of unmeasured or overestimated exclusive decays that include a $\phi$ meson.
A measurable difference between the inclusive and exclusive decay branching fractions
would indicate the size of as yet unmeasured decays, or would imply that some decays are overestimated, requiring complementary or more precise measurements. Previous measurements of the branching fractions for inclusive $D^+\to \phi X$ and $D^0\to \phi X$ decays were made by BES and CLEO~\cite{ref::BES, ref::CLEO-c} with 22.3 pb$^{-1}$ and 281 pb$^{-1}$ of $e^+e^-$ annihilation data samples taken at the center-of-mass energies $\sqrt s=$4.03 and 3.774 GeV, respectively.
Table~\ref{t_exclusive_phix} summarizes the branching fractions of the reported exclusive $D$ decays to $\phi$, where the branching fractions of $~D^+ \to \phi \pi^+$, $~D^0 \to \phi \pi^0$ and $~D^0 \to \phi \eta$ are quoted from the recent BESIII measurements~\cite{ref::phip}, the branching fraction of $D^+\to\phi K^+$ is from the LHCb measurements~\cite{ref::phik1, ref::phik2}, while the others are quoted from the Particle Data Group (PDG)~\cite{ref::PDG}. In this paper, we report improved measurements of the branching fractions of these inclusive decays by using 2.93 fb$^{-1}$ of $e^+e^-$ annihilation data taken at $\sqrt s=$3.773 GeV with the BESIII detector. Throughout this paper, the charged conjugate modes are implied
unless stated explicitly.

\renewcommand{\thefootnote}{\fnsymbol{footnote}}
\footnotetext{Corresponding author: zhaomg@nankai.edu.cn\\~~~~qusq@mail.nankai.edu.cn}

Secondly, charge-parity ($CP$) violation plays an important role in interpreting the matter-antimatter asymmetry in the Universe and in searching for new physics beyond the standard model (SM). It has been well established in the $K$- and $B$-meson systems.
In the SM, however, $CP$ violation in charm decays is expected to be much smaller~\cite{ref::CPV1, ref::CPV2, ref::CPV3}.
Searching for $CP$ violation in $D$ meson decays is important to explore physics beyond the SM. Recently, $CP$ violation in charm sector was observed for the first time in the charm hadrons decays at LHCb~\cite{ref::LHCb}.
In this paper, we search for $CP$ violation in the inclusive $D\to \phi X$ and $\bar D\to \phi X$ decays.

\begin{table}[b]
    \centering
    \caption{The branching fractions of the known exclusive decays  $D^{+(0)}\to\phi X$.}
    \label{t_exclusive_phix}
    \begin{tabular}{lr}
    \hline
    \hline
    \textbf{~Decay mode~} & \textbf{~$\mathcal{B}$~} \\
    \hline
    $~D^+ \to \phi \pi^+ \pi^0 $~           & ~$ (2.3 \pm 1.0) \%$~                               \\
    $~D^+ \to \phi \rho^+ $~                & ~$ <1.5 \%$~                                        \\
    $~D^+ \to \phi \pi^+$~    & ~$(5.70\pm0.14)\times 10^{-3}$~                            \\
    $~D^+ \to \phi K^+$~        & ~$(4.36\pm0.56)\times 10^{-6}$~                            \\
    \hline
    \textbf{~Sum~} & $(2.87\pm1.00)\%$\\
    \hline
    $~D^0\,\to \phi \gamma $~               & ~$ (2.81 \pm 0.19)\times 10^{-5}$~                      \\
    $~D^0\,\to \phi K_S^0$~ & ~$ (4.13\pm 0.31)\times 10^{-3}$~                       \\
    $~D^0\,\to \phi K_L^0$~ & ~$ (4.13\pm 0.31)\times 10^{-3}$~                       \\
    $~D^0\,\to \phi \omega $~               & ~$ < 2.1 \times 10^{-3}$~                               \\
    $~D^0\,\to \phi (\pi^+ \pi^-)_{S-\rm wave}$~&~$(20 \pm 10) \times 10^{-5}$~   \\
    $~D^0\,\to (\phi \rho^0)_{S-\rm wave}$~     &~$(14.0 \pm 1.2) \times 10^{-4}$~     \\
    $~D^0\,\to (\phi \rho^0)_{D-\rm wave}$~     &~$(8.5 \pm 2.8) \times 10^{-5}$~     \\
    $~D^0\,\to (\phi \rho^0)_{P-\rm wave}$~     &~$(8.1 \pm 3.8) \times 10^{-5}$~     \\
    $~D^0 \to \phi \pi^0$~ & ~$(1.17\pm0.04) \times 10^{-3}$~\\
    $~D^0 \to \phi \eta$~ & ~$(1.81\pm0.46) \times 10^{-4}$~\\
    \hline
    \textbf{~Sum~} & $(1.14\pm0.02)\%$\\
    \hline
    \hline
    \end{tabular}
\end{table}

\section{BESIII DETECTOR AND MONTE CARLO SIMULATION}
The BESIII detector is a magnetic
spectrometer~\cite{ref::BESIII} located at the Beijing Electron
Positron Collider (BEPCII)~\cite{Yu:IPAC2016-TUYA01}. The
cylindrical core of the BESIII detector consists of a helium-based
 multilayer drift chamber (MDC), a plastic scintillator time-of-flight
system (TOF), and a CsI(Tl) electromagnetic calorimeter (EMC),
which are all enclosed in a superconducting solenoidal magnet
providing a 1.0~T magnetic field. The solenoid is supported by an
octagonal flux-return yoke with resistive plate counter muon
identifier modules interleaved with steel. The acceptance of
charged particles and photons is 93\% over $4\pi$ solid angle. The
charged-particle momentum resolution at $1~{\rm GeV}/c$ is
$0.5\%$, and the $dE/dx$ resolution is $6\%$ for the electrons
from Bhabha scattering. The EMC measures photon energies with a
resolution of $2.5\%$ ($5\%$) at $1$~GeV in the barrel (end cap)
region. The time resolution of the TOF barrel part is 68~ps, while
that of the end cap part is 110~ps. The end cap TOF
system was upgraded in 2015 with multi-gap resistive plate chamber
technology, providing a time resolution of 60~ps~\cite{etof}. More details about the design and performance of the detector are given in Ref.~\cite{ref::BESIII}.

Simulated samples of events produced with the {\sc
geant4}-based~\cite{ref::GEANT4} Monte Carlo (MC) package, which
includes the geometric description of the BESIII detector and the
detector response, are used to determine the detection efficiency
and to estimate the backgrounds. The simulation includes the beam
energy spread and initial state radiation (ISR) in the $e^+e^-$
annihilations modeled with the generator {\sc
kkmc}~\cite{ref::KKMC1, ref::KKMC2}.
The inclusive MC samples consist of the production of $D\bar{D}$
pairs with consideration of quantum coherence for all neutral $D$
modes, the non-$D\bar{D}$ decays of the $\psi(3770)$, the ISR
production of the $J/\psi$ and $\psi(3686)$ states, and the
continuum processes incorporated in {\sc kkmc}~\cite{ref::KKMC1, ref::KKMC2}.
The known decay modes are modeled with {\sc
evtgen}~\cite{ref::EVTGEN1, ref::EVTGEN2} using branching fractions taken from the
Particle Data Group~\cite{ref::PDG}, and the remaining unknown charmonium decays are modeled by {\sc
lundcharm}~\cite{ref::LUNDCHARM}. Final state radiation
from charged final state particles is incorporated with the {\sc
photos} package~\cite{ref::PHOTONS}.

\section{Analysis method}
As the $\psi(3770)$ resonance peak lies just above $D\bar{D}$ threshold, it decays predominately into $D\bar{D}$ meson pairs. This advantage is leveraged by using a double-tag method, which was first developed by the MARKIII Collaboration~\cite{ref::MARK1, ref::MARK2}, to determined absolute branching fractions. If a $\bar D$ ($D^-$ or $\bar D^0$) meson is found in an event, the event is identified as a ``single-tag (ST) event". If the partner $D$ ($D^+$ or $D^0$) is reconstructed in the rest of the event, the event is identified as a ``double-tag (DT) event". In this analysis, the ST $D^-$ mesons are reconstructed by using $K^+\pi^-\pi^-$, $K^+\pi^-\pi^-\pi^0$, $K_{S}^{0}\pi^-$, $K_{S}^{0}\pi^-\pi^0$ and $K_{S}^{0}\pi^-\pi^-\pi^+$, and the ST $\bar{D^0}$ mesons are reconstructed by using $K^+\pi^-$, $K^+\pi^-\pi^0$ and $K^+\pi^-\pi^-\pi^+$. The signal $D^+$ and $D^0$ mesons are reconstructed by using $\phi X$, $\phi\to K^+K^-$. 
The branching fraction for $D\to\phi X$ decay is given by
\begin{linenomath}
\begin{equation}
\mathcal{B}_{\rm sig}=\frac{N_{\rm DT}}{\sum_{i}(N_{\rm ST}^{i}\cdot\epsilon_{\rm DT}^{i}/\epsilon_{\rm ST}^{i}/f_{\rm QC}^i})=\frac{N_{\rm DT}}{(N_{\rm ST}\cdot\epsilon_{\rm sig})},
\end{equation}
\end{linenomath}
where $i$ is the $i$-th ST mode, $N_{\rm DT}^i$ and $N_{\rm ST}^i$ are the yield of the DT and ST events, $\epsilon_{\rm ST}^{i}$ is the efficiency for reconstructing the tag candidate, and $\epsilon_{\rm DT}^{i}$ is the efficiency for simultaneously reconstructing the $\bar{D}$ decay to tag mode $i$ and $D$ decay to $\phi X$. $N_{\rm DT}$ and $N_{\rm ST}$ are the total yields of the DT and ST events, and $\epsilon_{\rm sig}=\sum_{i}(N_{\rm ST}^{i}\cdot\epsilon_{\rm DT}^{i}/\epsilon_{\rm ST}^{i}/f_{\rm QC}^i)/N_{\rm ST}$ is the average efficiency of finding the signal decay, weighted by the yields of tag modes in data. Here $f_{\rm QC}^i$ is a factor to take into account the quantum-correlation (QC) effect in $D^0\bar D^0$ pairs, called QC correction factor. The $f_{\rm QC}^i$ is taken as unity for charged $D$ tags, but determined for neutral $D$ tags 
following Refs.~\cite{ref::QC0,ref::QC00} (See Appendix for more details).

%

\section{Selection and yield of ST $\bar{D}$ mesons}
\label{sec:dtag}
All charged tracks, except those originating from $K_{S}^{0}$ decays, are required to originate in the interaction region, which is defined as $V_{xy}<1$ cm, $|V_z|<10$ cm, $|\rm cos\theta|<0.93$, where $V_{xy}$ and $|V_z|$ denote the distances of the closest approach of the reconstructed track to the Interaction Point (IP) perpendicular to and parallel to the beam direction, respectively, and $\theta$ is the polar angle with respect to the beam axis. Charged tracks are identified using confidence levels for the kaon (pion) hypothesis $CL_{K(\pi)}$~\cite{ref::BESIII}, calculated with both $dE/dx$ and TOF information. The kaon (pion) candidates are  required to satisfy $CL_{K(\pi)}>CL_{\pi(K)}$ and $CL_{K(\pi)}>0$.
 The $K_{S}^{0}$ candidates are formed from two oppositely charged tracks with $|V_z|<20$ cm and $|\rm cos\theta|<0.93$. The two charged tracks are assumed to be a $\pi^+\pi^-$ pair without Particle Identification (PID) and the $\pi^+\pi^-$ invariant mass must be within (0.487, 0.511) GeV/$c^2$.
 The photon candidates are selected from isolated EMC clusters. To suppress electronics noise and beam backgrounds, the clusters are required to have a start time within 700 ns after the event start time and have an opening angle greater than 10$^\circ$ with respect to the nearest extrapolated charged track. The energy of each EMC cluster is required to be larger than 25 MeV in the barrel region ($|\rm cos\theta|<0.8$) or 50 MeV in the end-cap region ($0.86<|\rm cos\theta|<0.92$). To select $\pi^0$ meson candidates, the $\gamma\gamma$ invariant mass is required to be within (0.115, 0.150) GeV/$c^2$. The momentum resolution of $\pi^0$ candidates is improved with a kinematic fit that constrains the $\gamma\gamma$ invariant mass to the $\pi^0$ nominal mass~\cite{ref::PDG}. 
For $\bar D^0\to K^+\pi^-$ candidates, backgrounds arising from cosmic rays and BhaBha scattering events are rejected with the same requirements as those described in Ref.~\cite{ref::COMSIC}.

Two variables, the energy difference $\Delta E \equiv E_{\bar D}-E_{\rm beam}$ and the beam-energy-constrained mass $M_{\rm BC}\equiv \sqrt{E^2_{\rm beam}/c^4-p^2_{\bar D}/c^2}$, are used to identify the ST $\bar{D}$ candidates. Here, $E_{\rm beam}$ is the beam energy and $E_{\bar D}(p_{\bar D})$ is the reconstructed energy (momentum) of the ST $\bar{D}$ candidates in the center-of-mass frame of the $e^+e^-$ system. For a given tag mode, if there are multiple candidates per charm per event, the one with the smallest value of $|\Delta E|$ is retained. Combinatorial backgrounds are suppressed by mode dependent $\Delta E$ requirements, as shown in Table ~\ref{tab:dtag}.

\begin{figure}[htpb]
  \includegraphics[width=\columnwidth]{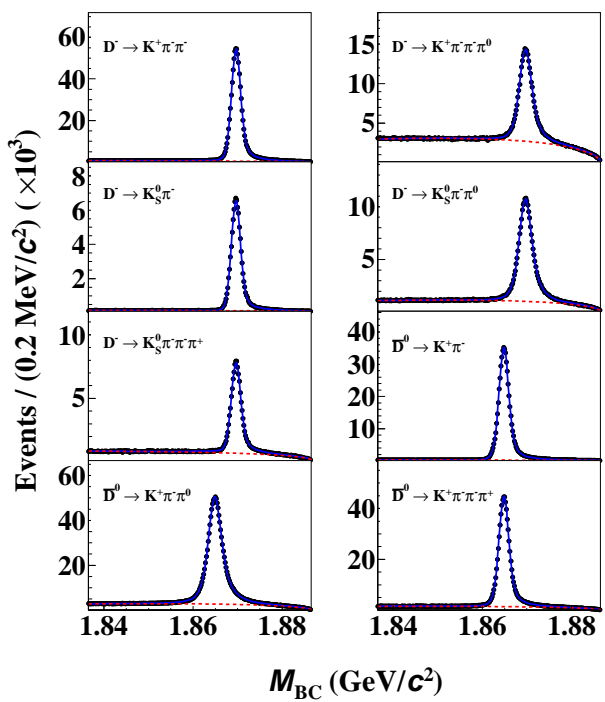}
  \caption{Fits to the $M_{\rm BC}$ distributions of the ST $\bar{D}$ meson candidates. The dots with error bars are data, the blue solid curves are the overall fits, and the red dashed curves are the fitted background shapes.}
  \label{fig:tag_mbc}
\end{figure}

Figure~\ref{fig:tag_mbc} shows the $M_{\rm BC}$ distributions of the accepted ST $\bar D$ candidates. The ST yields ($N_{\rm ST}^i$) for different tags are determined using a binned maximum likelihood fit to the corresponding $M_{\rm BC}$ distribution. An MC-simulated signal shape convolved with a double Gaussian function is used to model the $M_{\rm BC}$ signal and the combinatorial backgrounds in $ M_{\rm BC}$ distribution are modeled by an ARGUS function~\cite{ref::ARGUS} with the endpoint fixed at $E_{\rm beam}$. The ST efficiencies ($\epsilon_{\rm ST}^i$) are determined with inclusive MC samples. The ST yields in data within the $\Delta E$, $M_{\rm BC}$ signal regions and the corresponding ST efficiencies are summarized in Table~\ref{tab:dtag}.

\begin{table*}
  \caption{Summary of the $\Delta E$ requirements, the $M_{\rm BC}$ signal regions, the ST yields in data ($N_{\rm ST}^i$) and the ST efficiencies ($\epsilon_{\rm ST}^i$). The uncertainties are statistical only.}
  \label{tab:dtag}
  \begin{ruledtabular}
  \begin{tabular}{lcccc}
  Tag mode $i$ & $\Delta E$ (MeV) & $M_{\rm BC}$ (GeV/$c^2$) & $N_{\rm ST}^i$ & $\epsilon_{\rm ST}^i\,(\%)$\\
  \hline
  $D^-\to K^+\pi^-\pi^-$              & ($-$20, 19)  & (1.863, 1.879)  & $796040\pm1550$   & $50.70\pm0.04$\\
  $D^-\to K^+\pi^-\pi^-\pi^0$         & ($-$53, 30)  & (1.863, 1.879)  & $239070\pm737$    & $24.88\pm0.04$\\
  $D^-\to K_{S}^{0}\pi^-$             & ($-$23, 23)  & (1.863, 1.879)  & $93258\pm312$     & $51.52\pm0.12$\\
  $D^-\to K_{S}^{0}\pi^-\pi^0$        & ($-$61, 36)  & (1.863, 1.879)  & $204591\pm553$    & $27.13\pm0.08$\\
  $D^-\to K_{S}^{0}\pi^-\pi^-\pi^+$   & ($-$20, 18)  & (1.863, 1.879)  & $111994\pm1538$   & $27.82\pm0.15$\\
  \hline
  Sum                                 &            &                 & $1444953\pm2390$  &                     \\
  \hline \hline
  $\bar D^0\to K^+\pi^-$                   & ($-$25, 23)  & (1.858, 1.874)  & $537047\pm762$    & $66.00\pm0.06$\\
  $\bar D^0\to K^+\pi^-\pi^0$              & ($-$61, 36)  & (1.858, 1.874)  & $1075251\pm1415$  & $36.25\pm0.06$\\
  $\bar D^0\to K^+\pi^-\pi^-\pi^-$         & ($-$17, 15)  & (1.858, 1.874)  & $691228\pm952$    & $37.47\pm0.05$\\
  \hline
  Sum                                 &            &                 & $2303526\pm1867$  &                     \\
  \end{tabular}
  \end{ruledtabular}
\end{table*}

\section{Selection and yield of $D\to\phi X$}
\label{Dtophix}

DT events containing a $\phi$ meson are selected by investigating the system recoiling against the ST $D^-(\bar D^0)$.
Candidate DT events are required to have at least two good charged tracks with opposite charges.
The $\phi$ candidates are reconstructed through $\phi\to K^+K^-$ decays.
The selection and identification criteria of the charged kaons are identical to those for the tag side.

\begin{figure}
  \includegraphics[width=\columnwidth]{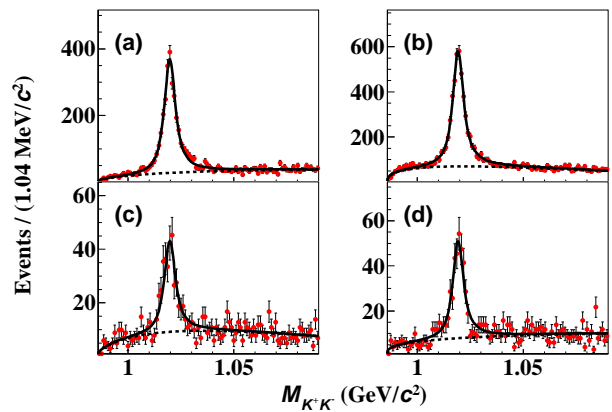}
  \caption{
Fits to the ${M}_{K^+K^-}$ spectra of the candidate events  for (a) $D^+\to\phi X$ and (b) $D^0\to\phi X$ in the $M_{\rm BC}$ signal region, (c) $D^+\to\phi X$ and (d) $D^0\to\phi X$ in the $M_{\rm BC}$ sideband region. The dots with error bars are data, the solid curves are the fit results,
and the dashed curves are the fitted combinatorial backgrounds.}
\label{fig:rec_yields}
\end{figure}

The $K^+K^-$ invariant mass (${M}_{K^+K^-}$) spectra of the accepted candidates for $D\to \phi X$ in the $M_{\rm BC}$ signal region are shown in the top row of Fig.~\ref{fig:rec_yields}. The events in the $M_{\rm BC}$ sideband region, (1.844, 1.860) GeV/$c^2$ for $D^+$ and (1.840, 1.856) GeV/$c^2$ for $D^0$, are used to estimate the peaking backgrounds in the ${M}_{K^+K^-}$ spectra, as shown in the bottom row of Fig.~\ref{fig:rec_yields}.
For each case, the yield of DT events containing $D\to\phi X$ signals is obtained by fitting these spectra. An MC-simulated signal shape convolved with a Gaussian function is used to model the $\phi$ signal and the combinatorial backgrounds are modeled by a reversed ARGUS background function~\cite{ref::ARGUS}.
The sideband contributions are normalized to the same background areas in the $M_{\rm BC}$ signal region.
The fit results are also shown in Fig.~\ref{fig:rec_yields}. The fitted DT yields in the $M_{\rm BC}$ signal and sideband regions in the data, $N_{\rm DT}^{{\rm sig}}$ and $N_{\rm DT}^{{\rm sid}}$, are given in Table~\ref{tab:rec_sum}.
The background-subtracted DT yields are calculated by $N^{\rm net}_{{\rm DT}}=N_{\rm DT}^{{\rm sig}}-f_{\rm co}N_{\rm DT}^{{\rm sid}}$, where
$f_{\rm co}$ is the ratio of the background area in the $M_{\rm BC}$ signal region over that in the $M_{\rm BC}$ sideband region
is determined to be 0.82 for the $D^+$ decay and 0.92 for the $D^0$ decay. These results have been verified by analyzing the inclusive MC sample.

\section{Branching fraction}

\begin{figure}[t]
  \includegraphics[width=\columnwidth]{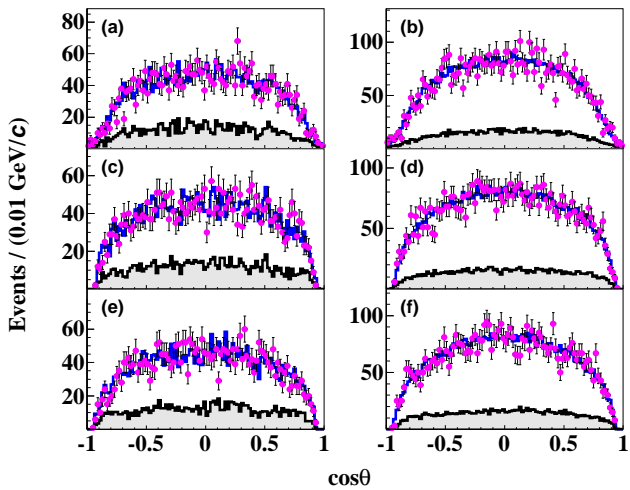}
\caption{
Comparisons of the $\cos\theta$ distributions for $\phi$ ((a) and (b)), $K^+$ ((c) and (d)), and $K^-$ ((e) and (f))
for the candidate events in $D^+\to\phi X$ and $D^0\to\phi X$, respectively.
The dots with error bars are the data, the solid histograms are the inclusive MC sample,
and the grey hatched histograms are the MC-simulated backgrounds.
An additional requirement of $|M_{K^+K^-}-1.019|<0.02$ GeV/$c^2$ has been imposed.
}
  \label{fig:compare_cos}
\end{figure}

\begin{figure}[t]
  \includegraphics[width=\columnwidth]{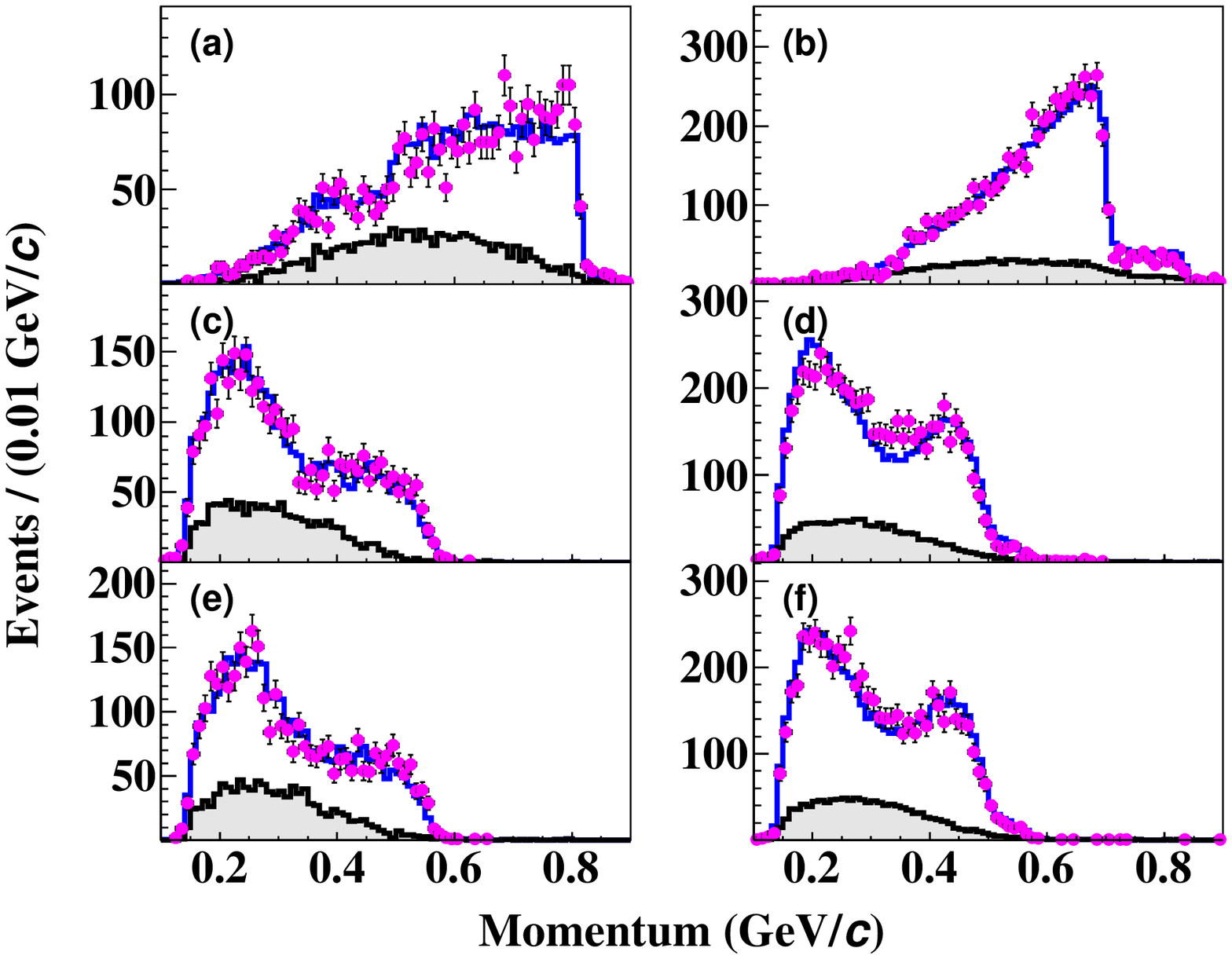}
  \caption{
Comparisons of the momentum distributions for $\phi$ ((a) and (b)), $K^+$ ((c) and (d)), and $K^-$ ((e) and (f))
for the candidate events in $D^+\to\phi X$ and $D^0\to\phi X$, respectively.
The dots with error bars are the data, the solid histograms are the inclusive MC sample,
and the grey hatched histograms are the MC-simulated backgrounds.
An additional requirement of $|M_{K^+K^-}-1.019|<0.02$ GeV/$c^2$ has been imposed.
}
  \label{fig:compare_p}
\end{figure}

The detection efficiencies are estimated by analyzing exclusive signal MC samples with the same procedure as for analyzing data. For the ST side, all possible sub-resonances have been included in the MC simulations. For the signal side, all known $D$ meson decays involving $\phi$ have been included in the MC simulations. Especially, to obtain better data/MC agreement, we have re-adjusted the branching fraction of $D^+\to\phi\pi^+\pi^0$, which is dominated by $D^+\to \phi\rho^+$, to be 0.6\% in the MC simulations. The efficiencies have been corrected by the small differences in $K^\pm$ tracking and PID between the data and MC simulation. To verify the reliability of the detection efficiencies, we compare the $\cos\theta$ and momentum distributions for $\phi$, $K^+$, and $K^-$ for the selected candidate events
in data and MC simulations, as shown in Figs. \ref{fig:compare_cos} and \ref{fig:compare_p}.
Good data-MC agreement is observed. The detection efficiencies and the measured branching fractions for $D\to\phi X$ are given in Table~\ref{tab:rec_sum}.

\begin{table*}
  \caption{
  Summary of the fitted DT yields in the $M_{\rm BC}$ signal and sideband regions ($N_{\rm DT}^{{\rm sig}}$ and $N_{\rm DT}^{{\rm sid}}$), background-subtracted DT yields ($N^{\rm net}_{{\rm DT}})$, signal efficiencies ($\epsilon_{\rm sig}$) and the measured branching fractions ($\mathcal{B}$). The uncertainties are statistical only.
 }
  \label{tab:rec_sum}
  \begin{ruledtabular}
  \begin{tabular}{lcccccc}
  Decay mode  & $N_{\rm ST}^{\rm tot}$ & $N_{\rm DT}^{\rm sig}$ & $N_{\rm DT}^{\rm sid}$ & $N_{\rm DT}^{\rm net}$ & $\epsilon_{\rm sig}\,(\%)$ & $\mathcal{B}\,(\%)$\\
  \hline
  $D^+\to\phi X$ & $721005\pm1673$  & $1478\pm50$ & $153\pm18$ & $1352\pm53$  & $16.69\pm0.20$ & $1.124\pm0.045$\\
  $D^-\to\phi X$ & $729840\pm1649$  & $1511\pm52$ & $155\pm18$ & $1384\pm55$  & $16.66\pm0.20$ & $1.141\pm0.046$\\
  $D^0\to\phi X$ & $1152037\pm1738$ & $2203\pm68$ & $185\pm19$ & $2033\pm70$  & $16.22\pm0.17$ & $1.088\pm0.037$\\
  $\bar{D^0}\to\phi X$ & $1146368\pm1529$ & $2239\pm66$ & $185\pm19$ & $2069\pm69$ & $16.46\pm0.17$ & $1.096\pm0.037$\\
  \hline
  $D^+/D^-\to\phi X$ & $1444953\pm2390$ & $2989\pm77$ & $302\pm25$ & $2741\pm81$  & $16.71\pm0.16$ & $1.135\pm0.034$\\
  $D^0/\bar{D^0}\to\phi X$ & $2303526\pm1867$ & $4441\pm98$ & $379\pm27$ & $4092\pm102$ & $16.28\pm0.13$ & $1.091\pm0.027$\\
  \end{tabular}
  \end{ruledtabular}
\end{table*}

Most of the systematic uncertainties originating from the ST selection criteria cancel when using the DT method. The systematic uncertainties in these measurements are assigned relative to the measured branching fractions and are discussed below.

The uncertainties due to the $M_{\rm BC}$ fits are estimated by using alternative signal shapes, varying the bin sizes, varying the fit ranges, and shifting the endpoint of the ARGUS background function.
We obtain 0.5\% as the total systematic uncertainty due to the $M_{\rm BC}$ fits.

The tracking and PID efficiencies for $K^{\pm}$ are studied by using DT $D\bar{D}$ hadronic events. In each case, the efficiency to reconstruct a kaon is determined by using the missing mass recoiling against the rest of the event and determining the fraction of events for which the missing kaon can be reconstructed. The differences in the momentum weighted efficiencies between the data and MC simulations (called the data-MC difference) due to tracking and PID are determined to be $(4.2\pm0.5)\%$ and $(0.5\pm0.5)\%$ per $K^\pm$.
After correcting the detection efficiencies obtained by MC simulations by these differences, the uncertainties of the data-MC differences are assigned as the systematic uncertainties for the $K^{\pm}$ tracking and PID efficiencies. This gives a systematic uncertainty for the $K^{\pm}$ tracking or PID efficiency of 0.5\% per track.

The systematic uncertainties arising from the fit range in the ${M}_{K^+K^-}$ fits are
estimated by a series of fits with alternative intervals. The maximum deviations in the resulting branching fractions are assigned as the associated systematic uncertainties, which are 0.4\% and 1.3\% for $D^+\to\phi X$ and $D^0\to\phi X$, respectively. To estimate the systematic uncertainties due to the signal shape in the ${M}_{K^+K^-}$ fits, we use a Breit Wigner function to describe the $\phi$ signal. The maximum deviations in the resulting branching fractions are assigned as the associated systematic uncertainties, which are 1.6\% and 1.8\% for $D^+\to\phi X$ and $D^0\to\phi X$, respectively. To estimate the systematic uncertainties due to the background shape in the ${M}_{K^+K^-}$ fits, we use an alternative background shape, $c_{1} \cdot (M_{K^+K^-}-M_\mathrm{threshold})^{1/2} + c_{3} \cdot (M_{K^+K^-}-M_\mathrm{threshold})^{3/2} + c_{5} \cdot (M_{K^+K^-}-M_\mathrm{threshold})^{5/2}$, to describe the background. The maximum deviations in the resulting branching fractions are assigned as the associated systematic uncertainties, which are 0.2\% and 1.6\% for $D^+\to\phi X$ and $D^0\to\phi X$, respectively. We assume that systematic uncertainties arising from the fit range, signal and background shape are independent and add them in quadrature to obtain the systematic uncertainty of the ${M}_{K^+K^-}$ fit.

In our nominal analysis, the measured branching fraction of $D^0\to \phi X$ has been corrected by an averaged QC factor $f_{\rm QC}$ defined in Sec.~VI. After this correction, we take the residual uncertainty of $f_{\rm QC}$, 0.5\%, as the systematic uncertainty due to the QC effect. The uncertainties due to limited MC samples are 0.8\% and 0.7\% for $D^+$ and $D^0$ decays, respectively. The uncertainty in the quoted branching fraction of $\phi\to K^+K^-$ is 1.0\%~\cite{ref::PDG}.

Assuming all the sources are independent, the quadratic sum of these uncertainties gives the total systematic uncertainty in the measurement of the branching fraction for each decay. Table~\ref{tab:systematic_uncertainty} summarizes the systematic uncertainties in the branching fraction measurements.

\begin{table}[htbp]
  \caption{Systematic uncertainties (in \%) in the measurements of the branching fractions.}
  \label{tab:systematic_uncertainty}
  \begin{ruledtabular}
  \begin{tabular}{lcc}
  Source & $D^+\to\phi X$ & $D^0\to\phi X$\\
  \hline
  $M_{\rm BC}$ fit       & 0.5 & 0.5 \\
  $K^{\pm}$ tracking     & 1.2 & 1.2 \\
  $K^{\pm}$ PID          & 1.0 & 1.0 \\
  ${M}_{K^+K^-}$ fit & 1.7 & 2.4 \\
  QC effect              & -   & 0.5 \\
  MC statistics          & 0.8 & 0.7 \\
  Quoted Branching Fraction             & 1.0 & 1.0 \\
  \hline
  Total                  & 2.7 & 3.2 \\
  \end{tabular}
  \end{ruledtabular}
\end{table}

\section{Asymmetry of ${\mathcal B}(D\to \phi X)$ and ${\mathcal B}(\bar D\to \phi X)$}

We determine the branching fractions of $D\to \phi X$ and $\bar D\to \phi X$ separately.
In this section, charge conjugated modes are not implied. Table \ref{tab:rec_sum} summarizes the ST yields,
the DT yields in the $M_{\rm BC}$ signal and sideband regions, detection efficiencies, and the measured branching fractions.
The asymmetry of the branching fractions of $D\to \phi X$ and $\bar D\to \phi X$ is determined by
\begin{linenomath}
\begin{equation}
{\mathcal A_{\rm CP}} = \frac{{\mathcal B}(D\to \phi X)-{\mathcal B}(\bar D\to \phi X)}{{\mathcal B}(D\to \phi X)+{\mathcal B}(\bar D\to \phi X)}.
\label{eq:cp}
\end{equation}
\end{linenomath}
The asymmetries for charged and neutral $D\to \phi X$ decays are determined to be $(-0.7\pm2.8\pm0.7)\%$ and  $(-0.4\pm2.5\pm0.7)\%$,
where the uncertainties due to the $M_{\rm BC}$ fit, $K^{\pm}$ tracking, $K^{\pm}$ PID, the ${M}_{K^+K^-}$ fit, the QC effect, and the quoted branching fractions in the measurements of ${\mathcal B}(D\to \phi X)$ and ${\mathcal B}(\bar D\to \phi X)$ cancel.
No CP violation is found at the current statistical and systematic precision.

\section{Summary}
By analyzing 2.93 fb$^{-1}$ of $e^+e^-$ annihilation data taken with the BESIII detector at $\sqrt{s}=3.773$ GeV, the branching fractions of $D^+\to\phi X$ and $D^0\to\phi X$ decays are measured to be $(1.135\pm0.034\pm0.031)\%$ and $(1.091\pm0.027\pm0.035)\%$, respectively, where the first uncertainties are statistical and the second systematic. Comparisons of our results with the previous measurements by CLEO~\cite{ref::CLEO-c} and BES~\cite{ref::BES} are shown in Table~\ref{comparison_results}. Our results are consistent with previous measurements, but with much better  precision.
These results indicate that the nominal values of the branching fractions for some known exclusive decays of the $D^+$ meson, $e.g.$, $D^+\to\phi\pi^+\pi^0$, may be overestimated. Precision measurements of some exclusive $\phi X$ decays of $D^+$ and $D^0$ mesons are required to further understand the discrepancy.
We also determine CP asymmetries in the branching fractions of $D\to\phi X$ and $\bar{D}\to\phi X$ decays for the first time, but no CP violation is found.

\begin{table*}[htpb]
  \caption{Comparisons of our branching fractions with the CLEO and BES results (\%).}
  \label{comparison_results}
  \begin{ruledtabular}
  \begin{tabular}{cccc}
      & This work & CLEO \cite{ref::CLEO-c}& BES \cite{ref::BES}\\
  \hline
  $D^+\to\phi X$ & $1.135\pm0.034\pm0.031$ & $1.03\pm0.10\pm0.07$ & $<1.8$\,(90\% C.L.) \\
  $D^0\to\phi X$ & $1.091\pm0.027\pm0.035$ & $1.05\pm0.08\pm0.07$ & $1.71^{+0.76}_{-0.71}\pm0.17$\\
  \end{tabular}
  \end{ruledtabular}
\end{table*}

\section{ACKNOWLEDGMENTS}
Authors thank for helpful discussions with Prof. Xueqian Li, Prof. Maozhi Yang and Dr. Haokai Sun.
The BESIII collaboration thanks the staff of BEPCII and the IHEP computing center for their strong support. This work is supported in part by National Key Basic Research Program of China under Contract No. 2015CB856700; National Natural Science Foundation of China (NSFC) under Contracts Nos. 11875170, 11775230, 11475090, 11335008, 11425524, 11625523, 11635010, 11735014; National Natural Science Foundation of China (NSFC) under Contract No. 11835012; the Chinese Academy of Sciences (CAS) Large-Scale Scientific Facility Program; Joint Large-Scale Scientific Facility Funds of the NSFC and CAS under Contracts Nos. U1832207, U1532257, U1532258, U1732263; CAS Key Research Program of Frontier Sciences under Contracts Nos. QYZDJ-SSW-SLH003, QYZDJ-SSW-SLH040; 100 Talents Program of CAS; INPAC and Shanghai Key Laboratory for Particle Physics and Cosmology; German Research Foundation DFG under Contracts Nos. Collaborative Research Center CRC 1044, FOR 2359; Istituto Nazionale di Fisica Nucleare, Italy; Koninklijke Nederlandse Akademie van Wetenschappen (KNAW) under Contract No. 530-4CDP03; Ministry of Development of Turkey under Contract No. DPT2006K-120470; National Science and Technology fund; The Knut and Alice Wallenberg Foundation (Sweden) under Contract No. 2016.0157; The Royal Society, UK under Contract No. DH160214; The Swedish Research Council; U. S. Department of Energy under Contracts Nos. DE-FG02-05ER41374, DE-SC-0010118, DE-SC-0012069; University of Groningen (RuG) and the Helmholtzzentrum fuer Schwerionenforschung GmbH (GSI), Darmstadt.

\begin{appendix}
\section{QC CORRECTION FACTOR}
At $\psi(3770)$, the $D^0\bar{D^0}$ pairs are produced coherently. The impact of the QC effect on the measurement of the branching fraction of $D^0\to\phi X$ is considered by two aspects: the strong-phase parameters of the tag modes and the $CP+$ fraction of the $D^0\to\phi X$ decay.

\subsection{Formulas}
Due to QC effect, the yield of the $i$-th ST candidates can be written as ~\cite{ref::QC0, ref::QC00}
\begin{linenomath}
\begin{equation}
\label{app:1}
N_{\rm ST}^i=(1+R_{{\rm WS}, f}^i)\cdot2N_{D^0\bar{D^0}}\cdot\mathcal{B}_{\rm ST}^i\cdot\epsilon_{\rm ST}^i,
\end{equation}
\end{linenomath}
and the yield of the DT candidates, i.e., $CP\pm$ eigenstate decay versus the $i$-th tag, can be written as
\begin{linenomath}
\begin{eqnarray}
\label{app:2}
N_{\rm DT}^i&=&(1+R_{{\rm WS}, f}^i\mp r_f^iz_f^i)\cdot2N_{D^0\bar{D^0}} \nonumber\\
&&\cdot\mathcal{B}_{\rm ST}^i\cdot\mathcal{B}_{\rm sig}^i\cdot\epsilon_{\rm DT}^i,
\end{eqnarray}
\end{linenomath}
where $N_{D^0\bar{D^0}}$ is the total number of $D^0\bar{D^0}$ pairs produced in data, $\epsilon_{\rm ST(DT)}^i$ is the efficiency of reconstructing the ST (DT) candidates, $\mathcal{B}_{\rm ST}^i$ and $\mathcal{B}^i_{\rm sig}$ are the branching fractions of the ST and signal decays, respectively, $R_{{\rm WS}, f}^i$ is the ratio of the Cabibbo-suppressed and Cabibbo-favored rates, $r_f^i$ is defined as $r^i_fe^{-i\delta^i_f}\equiv\frac{\langle f|\bar{D^0}\rangle}{\langle f|D^0\rangle}$, $z_f^i$ is defined as $z_f^i\equiv2{\rm cos}\delta_f^i$, $\delta_f^i$ is the strong-phase difference between these two amplitudes.

In this analysis, $R_{{\rm WS},f}^i$ is taken to be $r^2_i$, where $r_i$ is the ratio of the Cabibbo-suppressed and Cabibbo-favored amplitudes for $D^0\bar{D^0}$ decays to same final state. Then, we have
\begin{linenomath}
\begin{equation}
\label{app:3}
N^i_{\rm ST}=(1+r_i^2)\cdot2N_{D^0\bar{D^0}}\cdot\mathcal{B}_{\rm ST}^i\cdot\epsilon_{\rm ST}^i,
\end{equation}
\end{linenomath}

\begin{linenomath}
\begin{eqnarray}
\label{app:4}
N^i_{\rm DT}&=&(1+r_i^2\mp2r_iR_i{\rm cos}\delta_f^i)\cdot2N_{D^0\bar{D^0}} \nonumber\\
&&\cdot\mathcal{B}_{\rm ST}^i\cdot\mathcal{B}_{\rm sig}^i\cdot\epsilon_{\rm DT}^i,
\end{eqnarray}
\end{linenomath}
where $R_i$ is the coherence factor, $0<R_i\leq1$, that quantifies the dilution due to integrating over the phase space (For $D\to K^\pm\pi^\mp$, $R=1.00$)\cite{ref::QC11, ref::QC2}.

According to Eqs.~\ref{app:3} and ~\ref{app:4}, the absolute branching fraction for the signal decay is calculated by
\begin{linenomath}
\begin{equation}
\label{app:5}
\mathcal{B}_{\rm sig}^i=\frac{1}{1\mp C_f^i}\cdot\frac{N_{\rm DT}^i}{N_{\rm ST}^i\cdot(\epsilon_{\rm DT}^i/\epsilon_{\rm ST}^i)},
\end{equation}
\end{linenomath}
where $C_f^i$ is the strong-phase factor, which can be calculated by
\begin{linenomath}
\begin{equation}
\label{app:6}
C_f^i=\frac{2r_iR_i{\rm cos}\delta_f^i}{1+r_i^2}.
\end{equation}
\end{linenomath}

The amplitude of the neutral $D$ decays can be decomposed as mixture of the $CP+$ and $CP-$ components. This gives $F_+^{\rm sig}=1-F_-^{\rm sig}$, where $F_+^{\rm sig}$ and $F_-^{\rm sig}$ are the $CP+$ and $CP-$ fractions of the decay, respectively. The yield of the DT candidates tagged by the Cabibbo-favored tag mode $i$ can be written as
\begin{linenomath}
\begin{eqnarray}
\label{app:7}
N_{\rm DT}^i&=&F^{\rm sig}_+\cdot(1+r_i^2)\cdot(1-C_f^i)\cdot2N_{D^0\bar{D^0}} \nonumber\\
&&\cdot\mathcal{B}_{\rm ST}^i\cdot\mathcal{B}_{\rm sig}^i\cdot\epsilon_{\rm DT}^i+ \nonumber\\
&&F^{\rm sig}_-\cdot(1+r_i^2)\cdot(1+C_f^i)\cdot2N_{D^0\bar{D^0}} \nonumber\\
&&\cdot\mathcal{B}_{\rm ST}^i\cdot\mathcal{B}_{\rm sig}^i\cdot\epsilon_{\rm DT}^i. \nonumber\\
&=&[1-C_f^i\cdot(2F_+^{\rm sig}-1)]\cdot(1+r_i^2)\cdot2N_{D^0\bar{D^0}} \nonumber\\
&&\cdot\mathcal{B}_{\rm ST}^i\cdot\mathcal{B}_{\rm sig}^i\cdot\epsilon_{\rm DT}^i.
\end{eqnarray}
\end{linenomath}

According to Eqs.~\ref{app:3} and \ref{app:7}, the branching fraction of the signal decay can be calculated by

\begin{linenomath}
\begin{eqnarray}
\label{app:8}
\mathcal{B}_{\rm sig}^i&=&\frac{1}{1-C_f^i\cdot(2F_+^{\rm sig}-1)}\cdot\frac{N_{\rm DT}^i}{N_{\rm ST}^i\cdot(\epsilon_{\rm DT}^{{\color{blue}i}}/\epsilon_{\rm ST}^{{\color{blue}i}})} \nonumber\\
&=&f_{\rm QC}^i\cdot\frac{N_{\rm DT}^i}{N_{\rm ST}^i\cdot(\epsilon_{\rm DT}^{{\color{blue}i}}/\epsilon_{\rm ST}^{{\color{blue}i}})}.
\end{eqnarray}
\end{linenomath}
Here, $f_{\rm QC}^i=\frac{1}{1-C_f^i\cdot(2F_+^{\rm sig}-1)}$ is the QC correction factor to be determined.

\subsection{Strong-phase factor $C_f^i$}
Base on Eq.~\ref{app:6} and quoted parameters of $r_i$, $R_i$, and $\delta_f^i$, we obtain the strong-phase factor $C_f^i$ for the different ST modes. The quoted parameters of $r_i$, $R_i$ and $\delta_f^i$ as well as the obtained $C_f^i$ are listed in Table~\ref{app:tab1}.
\begin{table*}
  \centering
  \caption{Summary of the obtained $C_f$ and the parameters used to calculate the strong-phase factors.}
  \label{app:tab1}
  \begin{tabular}{lcccc}
  \hline
  \hline
  ~ST mode~ & ~$r$(\%)~ & ~$R$~ & ~$\delta_f (^\circ)$~ & ~$C_f$~\\
  \hline
  $D\to K^{\pm}\pi^{\mp}$                   & $5.86\pm0.02$~\cite{ref::QC1} & $1.00$ & $194.7_{-17}^{+8.4}$~\cite{ref::QC1} & $-0.113^{+0.004}_{-0.009}$\\
  $D\to K^{\pm}\pi^{\mp}\pi^0$              & $4.47\pm0.12$~\cite{ref::QC2} & $0.81\pm0.06$~\cite{ref::QC2} & $198.0^{+14}_{-15}$~\cite{ref::QC2} & $-0.069^{+0.008}_{-0.008}$\\
  $D\to K^{\pm}\pi^{\mp}\pi^{\mp}\pi^{\pm}$ & $5.49\pm0.06$~\cite{ref::QC2} & $0.43^{+0.17}_{-0.13}$~\cite{ref::QC2} & $128.0^{+28}_{-17}$~\cite{ref::QC2} & $-0.029^{+0.021}_{-0.014}$\\
  \hline
  \hline
  \end{tabular}
\end{table*}

\subsection{$CP+$ fraction of the signal decay}
According to Ref.~\cite{ref::QC3}, the $CP+$ fraction for the signal decay is determined by
\begin{linenomath}
\begin{equation}
\label{app:9}
F_+^{\rm sig}=\frac{N_+}{N_++N_-},
\end{equation}
\end{linenomath}
in which $N_{\pm}$ is the ratio of the DT and ST yields with $CP\mp$ tags, and is obtained by
\begin{linenomath}
\begin{equation}
\begin{split}
\label{app:10}
N_{\pm}=\frac{M_{\rm measured}^\pm}{S^\pm},\\
S^{\pm}=\frac{S_{\rm measured}^\pm}{1-\eta_\pm y_D},
\end{split}
\end{equation}
\end{linenomath}
where $M^{\pm}$ is the DT yields for $D^0\to\phi X$ versus $CP\mp$ tags, $S^{\pm}$ is the corrected ST yields for the $CP\pm$ decay modes. Here, $\eta_\pm=\pm1$ for $CP\pm$ decay modes and $y_D$ is the $D^0\bar{D^0}$ mixing parameter from the HFAG average~\cite{ref::PDG}.

\begin{figure}[b]
  \includegraphics[width=\columnwidth]{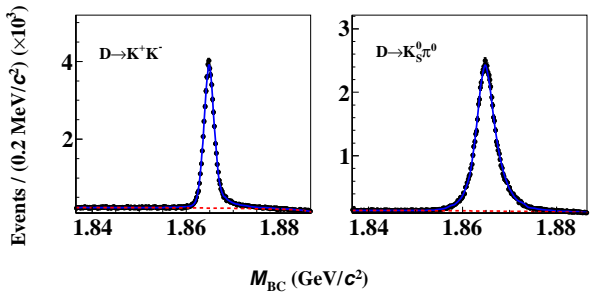}
  \caption{Fit to the $M_{\rm BC}$ distributions of the $D\to K^+K^-$ and $D\to K_S^0\pi^0$ candidates. The dots with error bars are data, the blue solid curves are the overall fits, and the red dashed curves are the fitted background shapes.}
  \label{fig:tag_mbc_cp}
\end{figure}

\begin{figure}[b]
  \includegraphics[width=\columnwidth]{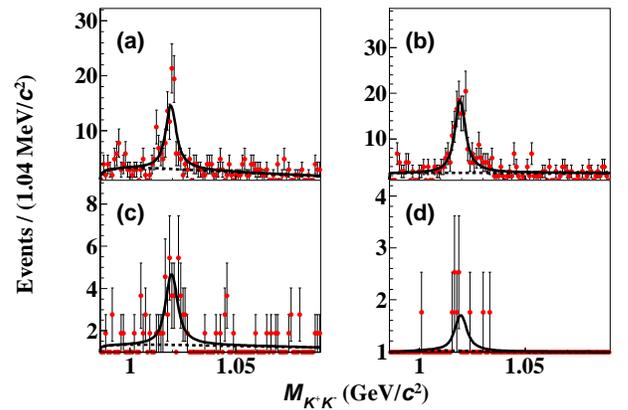}
  \caption{Fits to the $M_{K^+K^-}$ spectra of the candidate events for $D^0\to\phi X$ tagged by ((a) and (c)) $\bar{D^0}\to K^+K^-$ and ((b) and (d)) $\bar{D^0}\to K_S^0\pi^0$ in the $M_{\rm BC}$ signal and sideband regions, respectively. The dots with error bars are data, the solid curves are the fit results, and the dashed curves are the fitted combinatorial backgrounds.}
  \label{fig:signal_cp}
\end{figure}

To extract $F_+^{\rm sig}$ of the $D^0\to\phi X$ decay, we use the $CP+$ tag of $D\to K^+K^-$ and the $CP-$ tag of $D\to K_S^0\pi^0$. Figures~\ref{fig:tag_mbc_cp} and \ref{fig:signal_cp} show the fits to the $M_{\rm BC}$ distributions of the ST candidates and the $M_{K^+K^-}$ distributions of the DT candidates. From the fits, we obtain the measured ST and DT yields ($S^\pm_{\rm measured}$ and $M_{\rm measured}^{\pm}$), as summarized in Table~\ref{app:tab2}. Inserting these numbers in Eqs.~\ref{app:9} and \ref{app:10}, we obtain $F_+^{\rm sig}=0.64\pm0.05$.

\begin{table}
  \centering
  \caption{Summaries of the data yields and the MC efficiencies for the ST and DT candidates.}
  \label{app:tab2}
  \begin{tabular}{lcc}
  \hline
  \hline
  ~Decay mode~ & ~$D\to K^+K^-$~ & ~$D\to K_S^0\pi^0$~\\
  \hline
  ~$S_{\rm measured}^\pm$~  & ~$57147\pm372$~ & ~$65407\pm309$~ \\
  ~$M_{\rm measured}^\pm$~  & ~$73\pm15$~ & ~$147\pm15$~ \\
  \hline
  \hline
  \end{tabular}
\end{table}

\subsection{Impact on the measured branching fraction}
Inserting the $C_f^i$ and $F_+^{\rm sig}$ obtained above in Eqs.~\ref{app:6} and \ref{app:9}, we obtain the QC correction factors for the $D\to K^\pm\pi^\mp$, $D\to K^\pm\pi^\mp\pi^0$ and $D\to K^\pm\pi^\mp\pi^\mp\pi^\pm$ ST decays to be $(96.9\pm0.3\pm1.1)\%$, $(98.1\pm0.3\pm0.7)\%$ and $(99.2\pm0.7\pm0.3)\%$, where the first and second uncertainties are from $C_f^i$ and $F_+^{\rm sig}$, respectively.

\end{appendix}

\end{document}